\documentstyle[epsf,psfig]{mn}

\newcommand{\SS}{\scriptscriptstyle}
\newcommand{\Rl}{R_{\SS L1}}
\newcommand{\Md}{\dot{\cal M}}
\newcommand{\Msol}{{\cal M}_\odot}

\newcommand{\lsim}{{\textstyle{\; \lower 0.7ex\hbox{$<$}\;
  \atop \raise-0.1ex\hbox{$\sim$}}}}

\newcommand{\gs}{gs$^{-1}$}

\begin{document}

\title{V2051~Oph's disc evolution on decline from superoutburst}

\author[Sonja~Vrielmann and Warren~Offutt]
{Sonja~Vrielmann$^{1,2}$\thanks{Send offprint requests to: S.\ Vrielmann},
Warren Offutt$^3$\\
$^1$Department of Astronomy, University of Cape Town, Private Bag,
Rondebosch, 7700, South Africa (sonja@mensa.ast.uct.ac.za)\\
$^2$Hamburger Sternwarte, Universit\"at Hamburg, Gojenbergsweg 112,
	21029 Hamburg, Germany (svrielmann@hs.uni-hamburg.de)\\
$^3$W\&B Observatory, P.O. Drawer 1130, Cloudcroft, NM 88317, USA
(offutt@apo.nmsu.edu)\\
}

\maketitle

\begin{abstract}
We present an Eclipse Mapping analysis of ten eclipses taken during
decline from superoutburst of the dwarf nova V2051~Oph. On decline
from superoutburst the disc cools down considerably from nearly
$50,000$~K in the intermediate disc ($\sim 0.2 \Rl$) near maximum to
about 25,000~K at the end of our observing run, i.e.\ within 4 days. The
average mass accretion rate through the disc drops in the same time
from $10^{18}$~\gs\ to below $10^{17}$~\gs.

While in some maps the brightness temperature follows the steady state
model, in others the temperature profile shows flattenings and/or
indication of an inward travelling cooling front with a speed of
approximately $-0.12$~km~s$^{-1}$, possibly a reflected heating front
with a speed of $+1.8$~km~s$^{-1}$ and a newly reflected cooling front
with the same speed as the first one. Such scenario has been predicted
(Menou et al.\ 2000) but not been observed before. Furthermore, we see
a prograde precession of the enlarged disc with a precession period of
about 52.5$^h$ in very good agreement with the independently
determined superhump period observed by Kiyota \& Kato (1998). At the
same time, the uneclipsed component -- presumeably a disc wind --
decreases significantly in strength during decline from superoutburst.
\end{abstract}

\begin{keywords}
binaries: eclipsing -- novae, cataclysmic variables -- accretion,
accretion discs -- stars: V2051~Oph
\end{keywords}

\section{Introduction}
Cataclysmic variables (CVs) are close, interacting binaries in which a
Roche-lobe filling main sequence star (the secondary) looses matter to
its somewhat more massive white dwarf companion. In the absence of
strong magnetic fields the angular momentum conservations leads to the
formation of an accretion disc around the primary component through
which the matter from the secondary is accreted onto the compact
object. A detailed description of such systems is given in Warner
(1995), an introduction in Hellier (2001).

If the mass transfer rate is below a certain limit, these discs
undergo outbursts, during which the system brightens by a few
magnitudes. The cause of such outbursts is the accumulation of
material in the disc leading to a temperature increase and
subsequently ionization of hydrogen. The changed opacity leads to a
sudden further increase in temperature. The system is in
outburst. However, with such a low mass accretion rate as is typical
for a dwarf nova this state cannot be maintained. Subsequently, due to
a decrease in surface density and temperature the hydrogen eventually
recombines and the disc returns to its pre-outburst, quiescent
state. Osaki (1996) gives an overview on mechanisms and outburst
types.

Certain CVs additionally show so-called superoutbursts. These outburst
last longer than normal outbursts, but show a similar maximum
brightness. Additionally, they are associated with so-called
superhumps, short brightenings with a period slightly dissimilar to
the orbital period. The superoutbursts are usually associated with the
accretion disc's apsidal precession (see e.g.\ Patterson et al. 2002).

The eclipsing cataclysmic variable V2051~Oph was discovered by
Sanduleak (1972). Only few outbursts of the system have been reported
(Warner \& O'Donoghue 1987, Warner \& Cropper 1983, Bateson 1980) with
a maximal magnitude in B of about 13 in September 1980, possibly in
April 1982 and in May 1984. It is possible that outbursts were missed
due to the faintness of the system (Wenzel 1984). In May 1998 the
first superoutburst was detected (Kiyota \& Kato 1998).

We observed V2051~Oph during this superoutburst between maximum and
near quiescence.  This paper presents an analysis of these data using
the Eclipse Mapping technique invented by Horne (1985).

\section{The data}

\begin{figure}
\hspace*{1cm}
\psfig{file=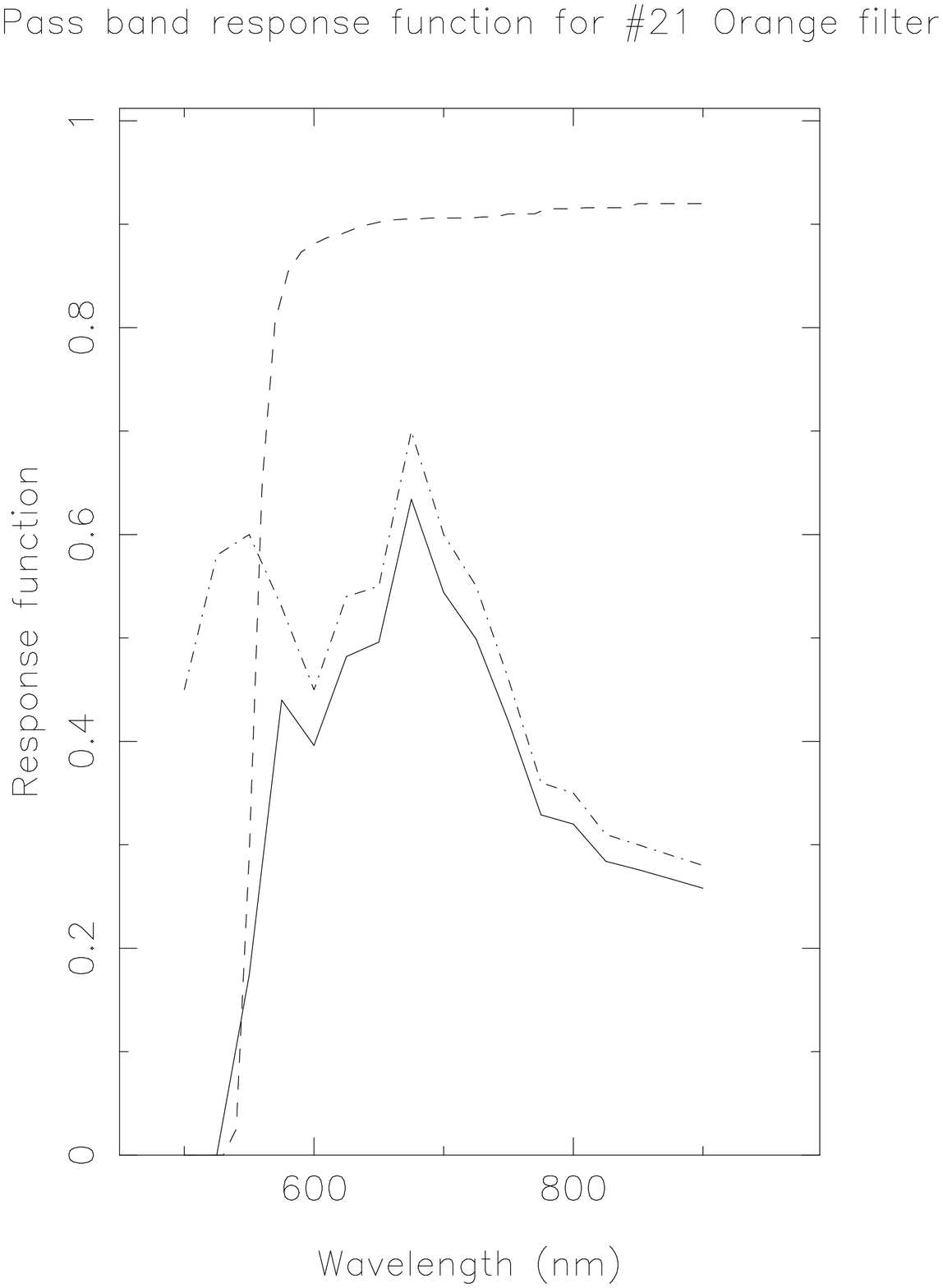,width=7.5cm}
\caption{\small The pass band response functions for the \#21 Orange
filter (dashed line) and the TC-241 CCD chip (dashed-dotted line) as
well as the product of both response functions (solid line).
\label{orange}}
\end{figure}

Our observing set consists of 10 eclipses taken in four nights between 23
and 27 May 1998 (interrupted by a cloudy night) during the
superoutburst first reported by R.\ Stubbings through VSNET. The data
were gathered by W.\ Offutt with the 60cm Ritchey-Chretien telescope
at the W\&B observatory in New Mexico/USA using a \#21 Orange filter
which covers R band and the long wavelength region of the V band, see
Fig.~\ref{orange}. For reasons described in Section~\ref{lightcurves} the
eclipse light curves were not averaged or phase binned, but left at
the original phase resolution of 10 to 15\,sec (about 0.0023 in phase).
A log of the superoutburst data is shown in Table~\ref{tab_sodata}.

\begin{table}
\caption{The superoutburst light curves of V2051~Oph.
\label{tab_sodata}}
\vspace{1ex}
\begin{tabular}{ccl}
\hspace{0.2cm}local evening date\hspace{0.2cm} & \hspace{0.3cm} HJD(start) \hspace{0.3cm} & Eclipse No.\\ \hline
23/5/98 & 2450957.24170 & S01, S02, S03\\
25/5/98 & 2450959.24344 & S33, S34\\
26/5/98 & 2450960.23838 & S49, S50, S51\\
27/5/98 & 2450961.22460 & S65, S66\\
\hline
\end{tabular}
\end{table}

According to VSNET observers the superoutburst started on May 17 when
the system rose to a magnitude of 13.5. It reached the maximum on May
18 with a magnitude of 11.7 and ended after May 27. Since this system
in quiescence is usually too faint for amateur astronomers it is not
possible to pin down the exact length of the outburst or even get a
complete superoutburst light curve.

Our series therefore started during decline of the superoutburst and
ended when the system got too faint for the telescope with the given
filter. This means we did not cover the full decline, but only down
to near the quiescent state.

For the phasing of the data we used the ephemeris of Echevarr\'{\i}a
\& Alvarez (1993). Since the white dwarf is most certainly outshown by
the accretion disc and due to the irregular eclipse profile (see
following Section) we did not attempt to determine a new ephemeris
from these superoutburst data.

\subsection{The light curves}
\label{lightcurves}
\begin{figure*}
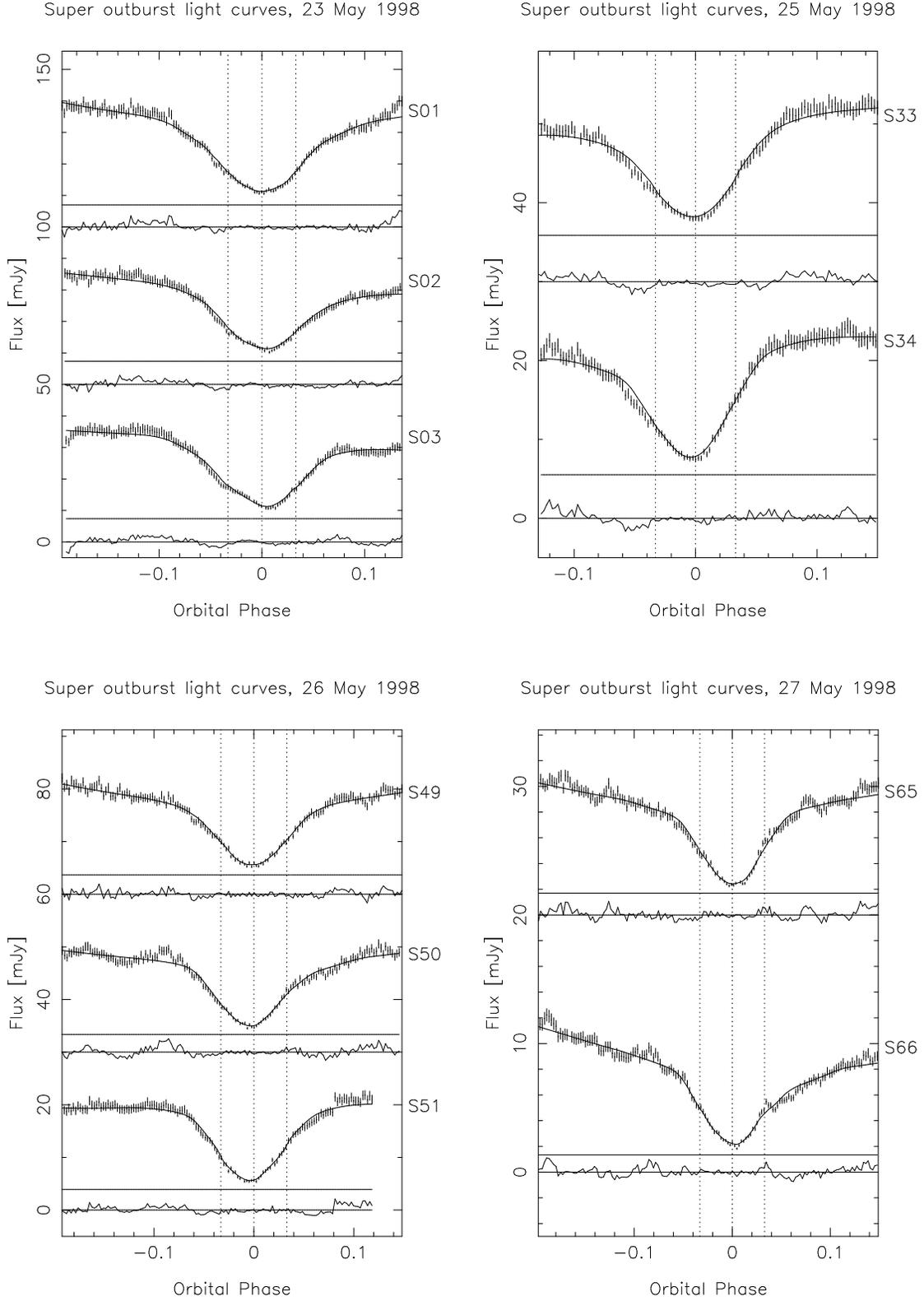

\hspace*{0.1cm}
\psfig{file=fig2a.ps,width=7cm}
\hspace*{0.5cm}
\psfig{file=fig2b.ps,width=7cm}

\vspace*{1cm}
\hspace*{0.1cm}
\psfig{file=fig2c.ps,width=7cm}
\hspace*{0.5cm}
\psfig{file=fig2d.ps,width=7cm}
\caption{\small The 10 superoutburst light curves together with
the PPEM fits. The eclipse light curve number is given on the right of
each panel.  Light curves are offset by 50mJy, 30mJy, 30mJy and 20mJy,
respectively, for the four different nights with horizontal lines
giving the zero-level for each light curve.  The residuals for each
light curve cross the zero-level lines.  Additional horizontal solid
lines indicate the reconstructed flux of the uneclipsed component (see
also Fig~\ref{fig_uc} and Table~\ref{tab_uc}).
Dashed vertical lines indicate phase 0 and the ingress and egress
phases of the white dwarf at -0.033 and 0.033 for comparision.
\label{lcv_outburst}}
\end{figure*}

Fig.~\ref{lcv_outburst} shows the observed eclipse light curves.  A
prelimimary analysis of the superoutburst data was shown in Vrielmann
\& Offutt 1999. But in that case we averaged the light curves of each
night, losing information about the short term variation of the disc.

As can be seen in Fig.~\ref{lcv_outburst}, the out-of-eclipse levels
and the eclipse profile change dramatically throughout the four
nights. The out-of-eclipse level drops from about 35~mJy down to
10~mJy and the eclipse minimum appears at positive or negative
phases. The shifts in eclipse minimum indicate that the disc
brightness distribution is asymmetric during superoutburst, with
occasionally the leading or the following lune of the disc being
brighter.

The eclipse shape is sometimes symmetric (as in eclipses S01,
S49, S50, S51 and S65) and in other cases very asymmetric with
out-of-eclipse levels before eclipse being higher (S02,S03,S66) or
lower (S33,S34). These general differences will appear in the
reconstructions as distinctive features.

\section{The analysis}
\subsection{Eclipse Mapping}

The eclipse mapping technique was invented by Horne (1985) as a
tomographic tool to reconstruct the intensity distribution in the
accretion disc by fitting the eclipse profile. Thus, intensity maps of
the accretion disc are derived using a maximum entropy algorithm (MEM,
Skilling \& Bryan 1984). The MEM algorithm serves to avoid the
ambiguities of reconstructing a two dimensional map from a
one-dimensional data set (the light curve) and introduces a certain
degree of symmetry in the map\footnote{For the current mapping we used
a solid arc smearing with smearing parameters 0.1$\Rl$ in radius and
20$^\circ$ in azimuth at the disc edge.}. Otherwise, a minimum of
assumptions about the system are made as to minimally influence the
results by premonitions.

The only assumptions are that the geometry of the system is known
(e.g. the secondary star fills its Roche-lobe, the accretion disc
is flat or of pre-determined 3-dimensional shape) and that the
accretion disc is not changing during the course of the eclipse as any
short term variation will lead to smeared out features in the
reconstructions. While the first assumption is generally relatively
well fullfilled, the latter assumption is usually not, since accretion
discs show short-term fluctuations in brightness called flickering. In
order to minimize this effect, usually a number of eclipse light
curves are averaged.

In the present case we did not average the eclipse light curves,
because we were interested in possible changes in the disc between
eclipses, e.g. due to a precession of the disc.  The precession may
cause a change of the orientation of the elliptical disc between two
consecutive eclipses. Therefore, we do not want to lose this
information by averaging over the eclipse light curves of a specific
night. This means, on the other hand, that we have to be careful not
to over-interpret structures in the light curve which are due to
flickering.

\subsection{The system parameters}
The inclination angle $i = 83.3^\circ$, the mass ratio $q = 0.19$, the
white dwarf mass ${\cal M}_{wd} = 0.78 \Msol$ and radius $R_{wd} =
0.0244 R_{\odot}$ as well as the scale parameter $\Rl/R_{\odot} =
0.42$ were taken from Baptista et al.\ (1998). For the distance we
used the Physical Parameter Eclipse Mapping estimate by Vrielmann,
Stiening \& Offutt (2002) of 146~pc.

\subsection{The disc geometry}
\label{discgeom}

In a first attempt to reconstruct the outburst disc we used a
geometrically infinitesimally thin disc in the orbital plane.
However, the disc appeared to be bright in the back part (disc region
opposite the seconary star) -- showing a so-called back-to-front
asymmetry --, indicating that the disc has a certain, non-negligible
thickness. This can also be seen in Z~Cha during outburst (Robinson,
Wood \& Wade 1999).  Therefore, for the study of the outburst disc we
use a 3-dimensional disc geometry.

We constructed a circular disc geometry with a fixed opening angle. At
the edge we allowed emission to be reconstructed on a vertical ribbon
around the disc. This geometrical model introduces two new parameters,
the opening angle and the disc radius (or disc size). However, it turned
out that the exact value of the opening angle did not influence the
results dramatically.  Similar results were found by Robinson et
al.~(1995) and Robinson (1999, private communication).

Since we used an inclination angle of $i=83.3^\circ$, we arbitrarily
chose a half opening angle of $\alpha=6.7^\circ$ (which allows us to
see just about all parts of the disc and the white dwarf) and a
somewhat smaller angle of $\alpha=4^\circ$. A comparison of the
reconstructions showed that in almost all cases the entropy
corresponding to the maps with the larger angle is somewhat larger
than the entropies for the thinner disc. In the remaining cases
(eclipses S01, S34, S51) the entropies of the reconstructions of both
angles were almost identical. Therefore we chose the larger angle of
$\alpha=6.7^\circ$.

Using a fixed opening angle means introducing an edge of the disc in the
form of a ribbon around the disc perimeter with 126 pixels onto which
intensity could be reconstructed as well as onto the circular disc
surface. This ribbon is set vertical to the orbital plane around the
disc at the disc edge $R_e$, therefore we can identify the ribbon pixel
uniquely by their azimuthal angle. We see at most half of this
ribbon, which half depending on the orbital phase. During eclipse, a
further part of this ribbon is occulted by the secondary. A similarly
flared disc geometry was used by Bobinger et al.~(1997).

Since the 3:1 resonance radius in V2051~Oph's accretion disc lies at
$R_{3:1} = 0.68\Rl$, the superoutburst disc is expected to be larger
than this. On the other hand, the Roche-lobe of the primary
confines the disc to within about $R_R = 0.8\Rl$. We chose a disc
edge at $R_e = 0.7\Rl$.

If the radius $R_e$ is chosen too small we should see this as
enhanced emission from the outer regions of the disc; if it is
too large, the reconstructions simply should not show much intensity in
the outer regions of the disc. Therefore, since the range between
$R_{3:1}$ and $R_R$ is small, the exact choice will not influence our
results dramatically. However, because of the MEM algorithm used in
Eclipse Mapping, the reconstructed brightness distribution in the disc
-- especially close to the disc edges -- is severely smeared out (see
also Appendix~\ref{app_test}) and will not be affected by a small
difference in true to assumed radius.

The introduction of a ribbon allowed us to use the light curves as
they were, without levelling the out-of-eclipse light curves. A flat
disc geometry cannot cope with different out-of-eclipse levels, but
in the flared disc approach intensity can be reconstructed onto the
ribbon. For example an orbital hump can be reconstructed on the leading
side of the ribbon, very well visible before eclipse and more or less
invisible after. However, in the present case we are dealing with
superoutburst data. We can therefore not distinguish between an
orbital hump and a superhump -- especially since our observations only
cover the immediate eclipses. Nevertheless, we did not attempt to
level out the out-of-eclipse light curves.

The use of a circular disc geometry seems to contradict the
expectation for a precession of the disc with an elongated,
elliptical disc shape. However, in order to avoid any assumption
about the direction of the semi major axis in the disc, we use a circular
disc geometry. In the case that the disc is truely elongated, we will
face similar scenarios as described above for a too small or too large
$R_e$, this time only applying to parts of the disc, i.e. we would
either see a lack of emission or enhancement close to the edge of two
regions $180^\circ$ apart from each other.

To identify the location of structures in the disc we use the radius
and azimuth. Azimuth $0^\circ$ indicates the direction from the white
dwarf to the secondary, negative azimuths correspond to the following
lune, positive angles to the leading lune of the accretion disc (the
side where the accretion stream hits the disc).

\begin{figure*}
\hspace*{0.1cm}
\psfig{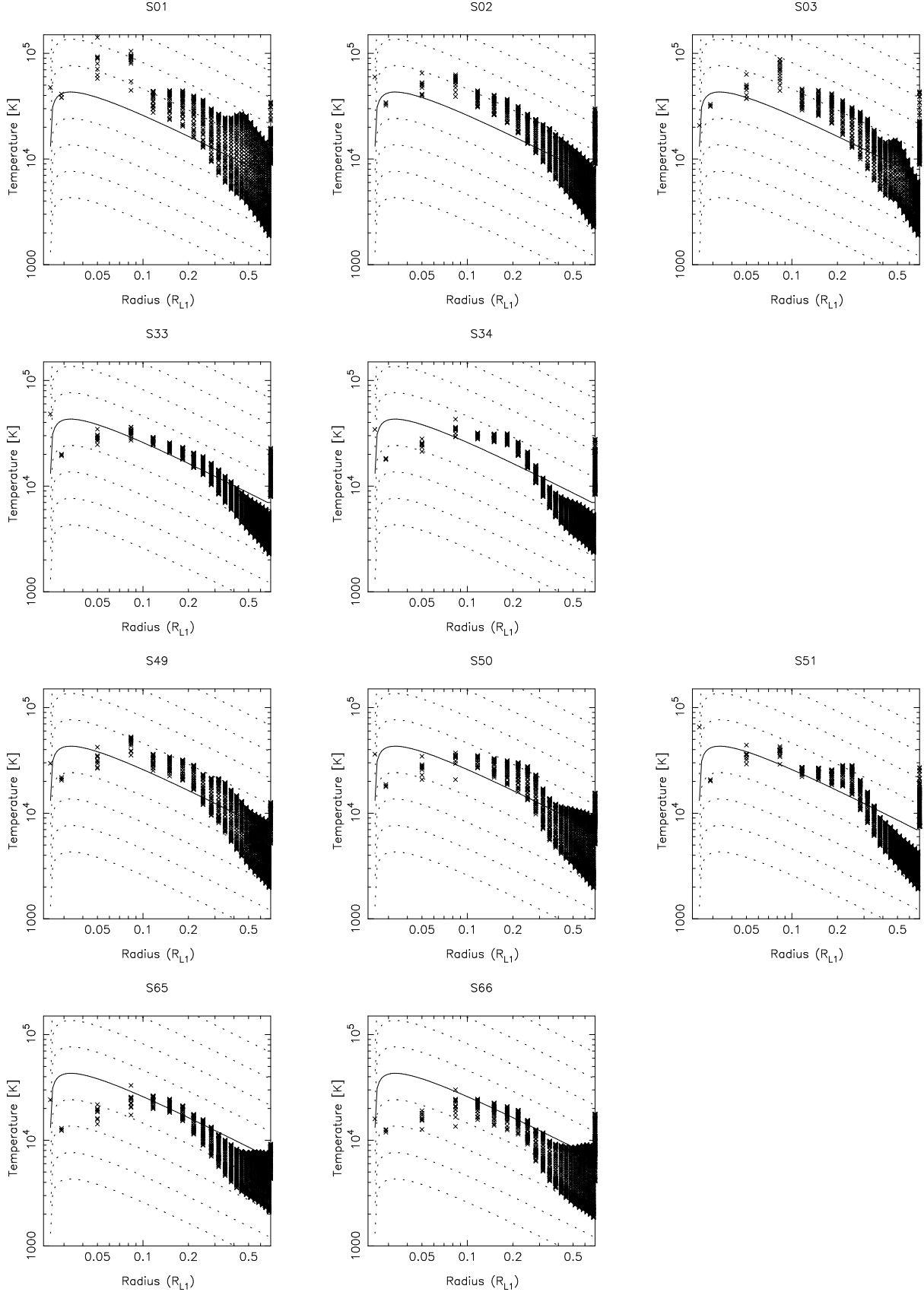}
\caption{\small The brightness temperatures of the four superoutburst
nights in the thick disc approximation. The number of 
the corresponding eclipse light curve is given above each panel.  (The
eclipses corresponding to the maps in the each column are separated by
multiples of almost exactly 24$^h$.) The panels show radial brightness
temperatures in the disc and the ribbon (at radius 0.7$\Rl$).
Underlying dashed lines are theoretical steady state distributions for
mass accretion $\log \Md = 13$ to 21, the one for $\Md = 10^{17}$\gs
is drawn solid for comparision of the four plots. Temperatures of
about 5000~K or below correspond to regions with no significant
emission. The scale is the same in all plots.
\label{temp_disc}}
\end{figure*}

\begin{figure*}
\hspace*{0.1cm}
\psfig{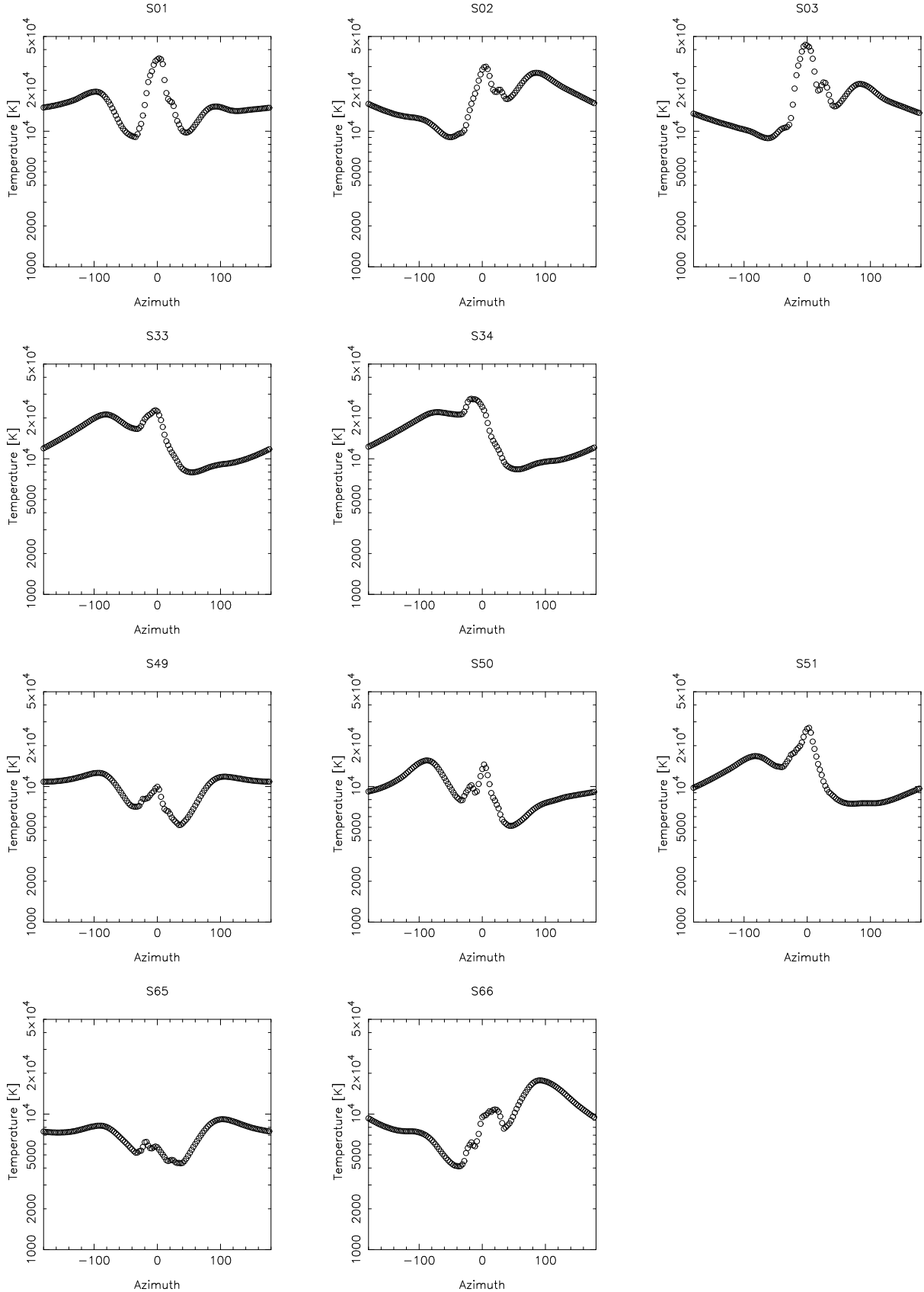}
\caption{\small The brightness temperatures on the ribbon during the
four superoutburst nights. The number of the corresponding
eclipse light curve is given above each panel.  The azimuth $0^\circ$
corresponds to the point closest to the secondary, $\pm 180^\circ$ to the
point farthest away from the secondary.  While a small scale variation
may be an artefact, one can distinguish a component with an apparent
movement from the leading side (positive angles) to the trailing side
(negative angles) of the disc during the first three
nights. Temperatures of about 5000~K or below correspond to regions
with no significant emission. The scale is the same in all plots.
\label{temp_rim}}
\end{figure*}

\begin{figure*}
\hspace*{0.1cm}
\psfig{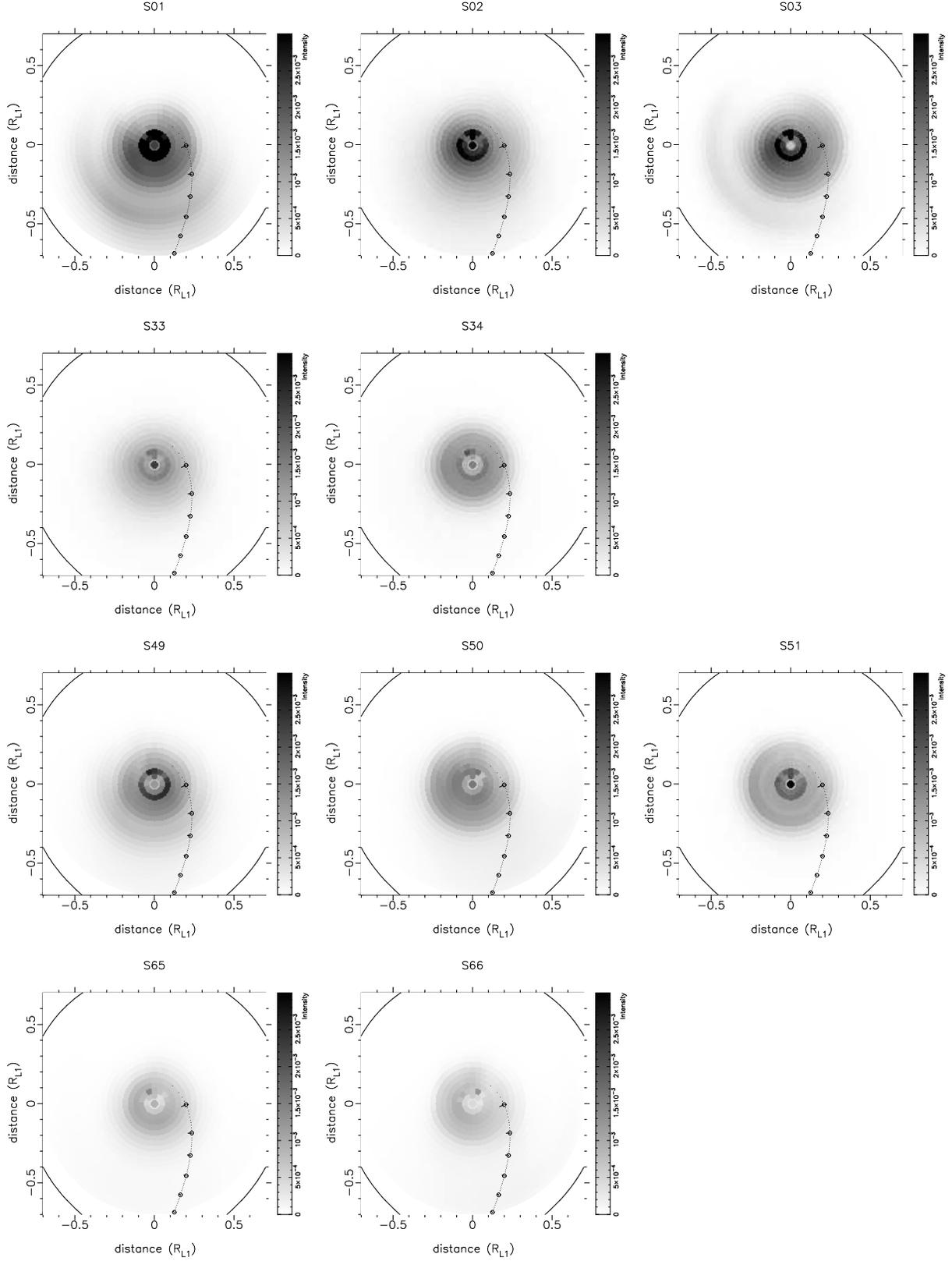}
\caption{\small The reconstructed intensity distributions for the
superoutburst disc as grey-scale plots. The number of the
corresponding eclipse light curve is given above each panel. Parts of
the Roche-lobe are plotted as a solid line and ballistic streams for a
mass ratio $q=0.19$ as dotted lines with the secondary positioned
below the plot, outside the plotted region. (The intensity in the ribbon is
not plotted, since it would not give more information than
Fig.~\ref{temp_rim}.) The scale is the same in all plots for
comparison. It is set to an ``overexposure'' in the maps of May
23rd to enhance the outer regions of the disc of the first night and
to allow the discs of the last night to be visible. (The maximum of
the intensity scale is lower than the maximum in the intensity values
of the maps in the first night.)
\label{map_rim}}
\end{figure*}

\subsection{The maps}

The Eclipse Mapping fits to the light curves (Fig.~\ref{lcv_outburst})
are good, we reached a $\chi^2$ of 1 for all light curves except for
light curves S50, S65 and S66 where we aimed only at a $\chi^2$ of 1.5
instead of 1. Aiming at a lower $\chi^2$ would have led to artificial
structures and noise in the maps.

Since the outburst disc is usually optically thick, the reconstructed
brightness distribution can directly be translated into a map of the
brightness temperature and treated as good approximations of the
true disc temperatures.

The disc reconstructions are displayed in Fig.~\ref{temp_disc}.  The
inner disc appears to be very hot on the first night, with
temperatures up to 100\,000~K in the disc centre and around 50\,000~K
in the intermediate disc ($\sim 0.2\Rl$). The temperature drops in the
following nights to values around 30\,000~K on May 25 and 26 and
20\,000~K in the last night within a radius of about $0.2\Rl$. These
values are compatible with theoretical considerations (Smak 2000,
Menou, Hameury \& Stehle 1999).

In the intermediate disc the temperature either follows a steady state
distribution (e.g.\ S02, S33, S49, S66) or is flat (S34, S51) or
inbetween. The mass accretion rate in the former cases is about
$10^{18}$ \gs\ in S02, $3 \times 10^{17}$ \gs\ in S33, $5 \times
10^{17}$ \gs\ in S49 and $10^{17}$ \gs\ in S66. The last value is
about a magnitude above the quiescence value of $10^{16}$ \gs\
(Vrielmann et al.\ 2002).

Fig.~\ref{temp_rim} shows the temperature distribution along the
ribbon. A comparison with Fig.~\ref{temp_disc} shows that the
reconstructed temperature in this ribbon is significantly higher than
in the disc edge and in some cases comparable to the temperature at a
radius of about $0.2\Rl$. As such feature does not appear in the test
cases, it is unlikely to be a systematic effect caused by the
method. It is possible that this indicates that the opening angle of
the disc is chosen too small, thus concentrating the emission on a too
small surface at the disc edge. Furthermore, this would mean that
parts of the disc surface are eclipsed by the high disc edge. As the
jump is strongest in those eclipses which show the strongest asymmetry
(i.e.\ S02, S03, S33, S34, with the exception of S51) this might
indicate an uneven height of the disc edge and a variablility of the
flaring angle during the course of the decline.

While for each night the distribution is similar, large
differences appear for different nights, reflecting a slow but
significant change of the disc. In Section~\ref{precession}
we describe what we can learn from these plots.

The grey-scale plots in Fig.~\ref{map_rim} give some indication of the
emissivity and possibly shape of the disc. However, since the MEM
algorithm introduces a certain amount of radial and azimuthal smearing
(and therefore a certain amount of symmetry) and due to a fixed
circular geometry we will not necessarily be able to directly
reconstruct an elliptical disc. A comparison to the test case in
Appendix~\ref{app_test}, however, shows that in all maps we can see an
asymmetry possibly indicating an elliptical -- definately non-circular
-- intensity distribution. The fact that in the maps the ``front''
instead of the ``back'' part is bright might indicate that the
opening angle we chose is too large, however, the bright region is not
always symmetric around azimuth $0^\circ$.

An interesting feature is the arc like structure in S01 and S03 and
possibly S49 (merged with the central part of the disc). Similar
structures have been observed in IP~Peg during decline from outburst
(Vrielmann 1997).

\subsection{The uneclipsed component}

\begin{figure}
\hspace*{0.1cm}
\psfig{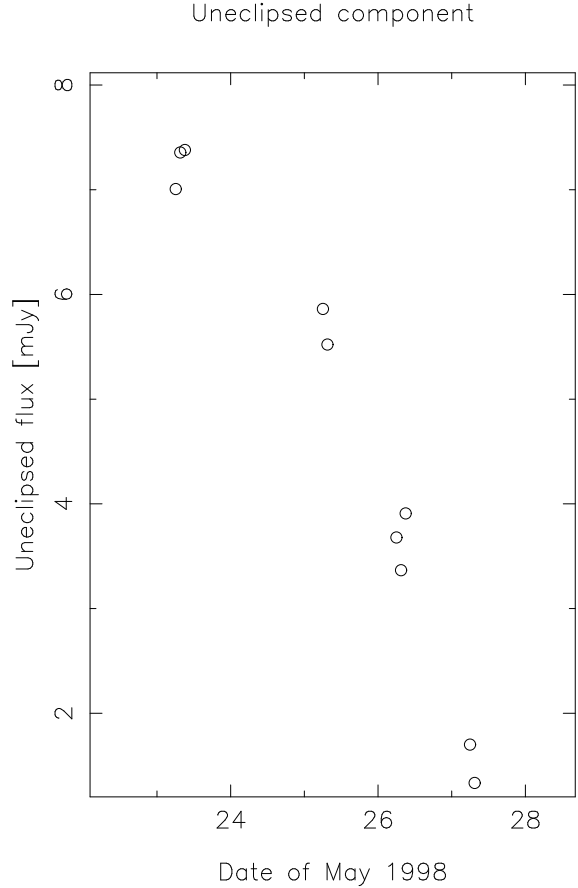}
\caption{\small The uneclipsed component during the four superoutburst
nights (cf.\ Table~\ref{tab_uc}).\label{fig_uc}}
\end{figure}

Table~\ref{tab_uc} and Fig.~\ref{fig_uc} give the reconstructed
uneclipsed component in the \#21 Orange filter for the observed
outburst nights. The scatter of the uneclipsed flux values in each
night is an indication for the error of these values of roughly half a
mJy. The flux level drops dramatically during the four nights.  Since
the filters R and \#21 Orange are similar, one can compare this flux
with the uneclipsed quiescent flux in the R band of 0.66 mJy (see
Vrielmann et al.\ 2002). This indicates that even in the fourth night
the uneclipsed light was not yet at its quiescent level.

\begin{table}
\caption{The reconstructed flux for the uneclipsed component in the
\#21 Orange filter for the four superoutburst nights.\label{tab_uc}}
\vspace{1ex}
\hspace{0.5cm}
\begin{tabular}{cccc}
Eclipse   & Night        & Time of mid-eclipse   &  Flux \\
no.       & (Date of May) & (in fractions of HJDs) & (mJy) \\ \hline
S01 & 23  &  .2538   &  7.01  \\
S02 &     &  .3162   &  7.36  \\
S03 &     &  .3787   &  7.38  \\
S33 & 25  &  .2515   &  5.86  \\
S34 &     &  .3139   &  5.52  \\
S49 & 26  &  .2503   &  3.68  \\
S50 &     &  .3128   &  3.37  \\
S51 &     &  .3752   &  3.91  \\
S65 & 27  &  .2492   &  1.70  \\
S66 &     &  .3116   &  1.33  \\
\hline
\end{tabular}
\end{table}

\section{Discussion}
\label{discussion}

\subsection{The disc evolution during superoutburst}

Assuming that in superoutburst the disc emits as an optically thick
source, we derive brightness temperatures from observations in an
orange filter (part of V and up to I). The reconstructed maps show that
the disc cools down considerably during the four nights.

Only in the intermediate eclipses (i.e.\ S34, S49, S50, S51, S66) do
we see indications for a cooling front as expected by theoretical
calculations (e.g.\ Cannizzo 1993).  Measuring the location of the
cooling front in S34 and S49 we derive a speed of $-0.12$~km~s$^{-1}$,
measuring the speed of the cooling front between S51 and S66 we
also derive a speed of about $-0.12$~km~s$^{-1}$, whereas these two
cooling fronts are not identical.

These speeds are comparable to the value Menou et al.\ (1999) found
with numerical simulations and who predict a fraction of a
km~s$^{-1}$. Furthermore, it is comparable with the speeds derived
with Eclipse Mapping of $-0.43$ to $-0.06$~km~s$^{-1}$ for EX~Dra
(decreasing during outburst, Baptista \& Catal\`an 2001), of
$-0.8$~km~s$^{-1}$ for IP~Peg (Bobinger et al.\ 1997) and $-0.7$ to
$-0.14$~km~s$^{-1}$ for OY Car in superoutburst (Bruch, Beele \&
Baptista 1996).

Now, why do we see two distinctly different cooling fronts? The
explanation can be found in numerical calculations by e.g.\ Menou et
al.\ (2000) who find that an inward moving cooling front can reflect
an outward moving heating front which again reflects into an inward
moving cooling front. If we assume this scenario we derive a speed of
$+1.8$~km~s$^{-1}$ for the reflected heating front.  A speed of a few
km~s$^{-1}$ is typical for heating fronts (Menou et al.\
1999). Furthermore, the reflection of the heating front occurs at a
disc radius $R_{front}$ of about $4\times10^{9}$~cm, compatible with Menou
et al.'s (2000) finding that they occur at $R_{front} \lsim
10^{10}$~cm in their simulations.

If such multiple reflections of cooling and heating fronts do occur in
V2051~Oph, we should observe re-flares as predicted by numerical
simulations (Menou et al.\ 2000). Unfortunately, the coverage of this
object by amateur astronomers (VSNET, AAVSO) during the so far
observed superoutbursts is too sparse to confirm or reject the
appearance of re-flares. Detailed observations during the next super
outburst are highly desirable.


\subsection{The disc precession}
\label{precession}

During superoutburst, we see a partially strong, variable peak in the
intensity distribution on the edge of the disc. It is probably caused
by an elliptical, precessing disc: Whenever one of the vertices of the
semimajor axis of the disc is close to the secondary, the disc will
experience tidal forces leading to precession and local heating of the
disc. The disc will react on the temperature increase by flaring up
near the secondary.  So, the location of the peak gives a rough
estimate for the phase of the apsidal precession. In
Appendix~\ref{app_test} we show that the location of a bright area
(not to be confused with the bright spot) on the disc edge can be
relatively well determined.

Looking at the distribution of light along the rim of the disc, we
identify three structures present in most reconstructions: (a) a
central peak close to azimuth $0^\circ$ (facing the secondary); (b) a
secondary peak in the profile of the central peak; and (c) a broad
underlying brightness distribution with a maximum at 90$^\circ$ and/or
-90$^\circ$. The latter peaks at the same side of the disc as the
secondary peak is located. Similar features appear in the test
(Appendix~\ref{app_test}). 

The central peak is also present in the test cases and therefore most
likely an artefact, although it could also be associated with the
bright spot where the ballistic stream hits the accretion disc. Such a
spot was not considered in the test case. The location of the bright
spot is only marginally influenced by a apsidal motion of an
elliptical, precessing disc. As apparent from the tests, the large
scale variation on the disc rim is most likely an artefact.

If we measure the position of the second peak, we find a 10$^\circ$
prograde shift per eclipse or a 10-20$^\circ$ backwards shift per
day. If this can be considered an indication for the apsidal motion of
the disc, we arrive at a value for the precession period $P_{prec}$ of
51$^h$ to 54$^h$, slightly larger than two days (therefore the
apparent backward shift during consecutive nights). This would mean
the disc is slightly larger than the 3:1 resonance radius $R_{3:1}$ of
0.68$\Rl$ and the superhump period ($P_{sh} = P_{prec} P_{orb}/
(P_{prec} - P_{orb}$), Warner 1995) should be 2.5 min or 3\% larger
than the orbital period $P_{orb} = 1.4982$ hours,
i.e. $P_{sh}$(predicted) = 1.5422$^h \pm$ 0.0013$^h$. This is in very
good agreement with the superhump period of 0.06423$^d$ = 1.54152$^h$
independently determined by Kiyota \& Kato (1998) during the same super
outburst.

This value for the precession period is very typical for SU~UMa type
dwarf novae, e.g.\ DV~UMa (67$^h$, Patterson et al.\ 2000), HT~Cas
(52$^h$), Z~Cha (51$^h$), OY~Car (65$^h$), IR Gem (47$^h$), AQ Eri
(52$^h$) to name but a few or even the nova V1974 Cyg with 44$^h$
(Retter, Leibowitz \& Ofek 1997). Numerical calculations of super
outbursts of dwarf novae with a similar mass ratio as V2051~Oph give
on average a slightly larger ratio of $P_{sh}/P_{orb}$ of about 1.06
(Murray 1998) compared to our 1.03.

\subsection{The mass accretion rate}

The mass accretion rate $\Md$ in superoutburst appears very high,
around $10^{18}$~\gs\ in our first night and varying throughout the
disc. This is about two magnitudes larger than in quiescence
(about $10^{16}$~\gs, Vrielmann et al.\ 2002). Towards the end of our
observing run (somewhat before the end of the outburst) the disc
cools down and shows a mass accretion rate of about $10^{17}$~\gs,
still a magnitude higher than the quiescent level.

Only occasionally the disc appears to follow the steady state models
and then only in intermediate disc regions between 0.1$\Rl$ and
0.2$\Rl$. At other times the distribution is rather flat. During the
last eclipse, when the disc is near the quiescent level, indication
for a bright spot can be seen at the expected location where the
accretion stream hits the disc.

Since the disc undergoes dramatic changes during the course of an
outburst with cooling fronts travelling through the material, we would
not expect it to settle down completely into a steady state. However,
the fact that parts of the disc do seem to appear in steady state
suggests that the time scale for returning into the steady state is
relatively short, of the order of a day or less.

According to Tylenda's (1981) calculations, the mass accretion rate in
superoutburst is too high and the disc is too small (outer radius at
$2.3 \cdot 10^{10}$ cm) to show optically thin areas -- the disc
should be everywhere optically thick. This agrees very well with our
assumption to consider the brightness temperatures as good
approximations of the gas temperatures.

\subsection{The disc wind}

The Eclipse Mapping shows that the uneclipsed component is decreasing
in strength on decline from outburst, but never disappearing
completely, as the analysis of the quiescence data show (Vrielmann et
al.\ 2002). A similar variability has been found for EX Dra by
Baptista \& Catal\'an (1999) where the out-of-eclipse flux in R dropped
during decline from outburst and practically disappeared at minimum
light.  At that time EX~Dra's disc is reduced to the immediate
vicinity of the white dwarf. Baptista \& Catal\'an suggest this variable
part of the uneclipsed component is due to a disc wind and a
chromosphere ejected by the inner part of the disc. The variation in
these components is presumably due to a variable mass accretion rate.

Our geometrical model of the superoutburst disc is very simple; the
disc may in fact be thicker, tilted, warped or only flare close to the
secondary. The latter geometrical model would introduce too many free
parameters (extent and location of flared part) which we cannot
determine from our data. However, the uneclipsed component is a flux
independent of orbital phase and therefore unlikely to originate in
the (eclipsed parts of the) disc itself, but rather in a region above
the disc. The easiest explanation is the same as for EX~Dra: a
variable disc wind triggered by the outburst.

\section*{acknowledgments}
We thank all amateur astronomers who are dedicating much of their
spare time to continously monitor dwarf novae in order to detect
outbursts. SV is funded by the South African Claude Harris Leon
Foundation through a postdoctoral fellowship.

\begin{appendix}
\section{Test for reconstructing elliptical discs}
\label{app_test}

\begin{figure*}
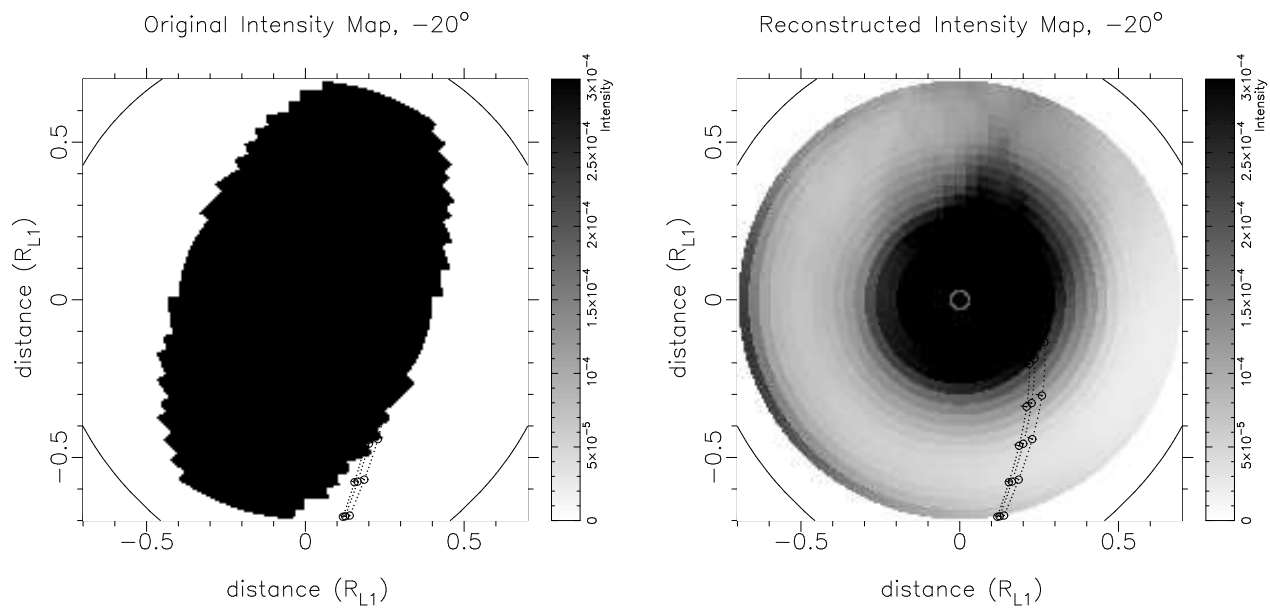

\hspace*{0.1cm}
\psfig{file=figA1a.ps,width=8cm}
\hspace*{0.5cm}
\psfig{file=figA1b.ps,width=8cm}

\caption{\small The original ({\em left}) and reconstructed ({\em
right}) maps for the test case in which the accretion disc shows an
elliptical intensity distribution on a circular disc geometry and has
an orientation angle of $-20^\circ$ in azimuth. The plots show
the intensity map on the surface of the accretion disc. The solid line
marks the Roche-lobe with the secondary being at the bottom, outside
the plotted area; the dotted line indicates the theoretical mass
accretion stream paths.
\label{app_plotv}}
\end{figure*}

In order to test how well an elliptical accretion disc can be
reconstructed, we performed a test in which we constructed an
artificial elliptical disc with an azimuthal angle of the semimajor
axis of $-20^\circ$ and a bright area at the vertex of the semimajor axis
closest to the secondary (see Fig.~\ref{app_plotv} ({\em left})). This
bright area is a simple simulation of the flaring of the disc near the
secondary. The geometry of the disc surface is the same as that
described in Section~\ref{discgeom}.

\begin{figure}
\hspace*{0.1cm}
\psfig{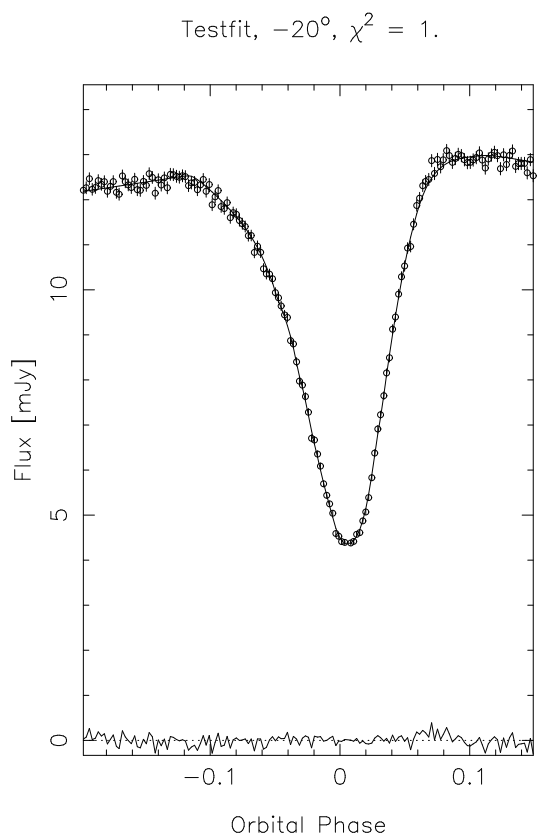}
\caption{\small The light curve constructed from the artificial disc
with the Eclipse Mapping fit (solid line through the data points) and
the residuals (solid line around zero Flux level).
\label{app_plotf}}
\end{figure}

For this artificial disc we calculated the light curve (see
Fig.~\ref{app_plotf}) that is to be fitted with the Eclipse Mapping
algorithm. Artificial noise with a signal-to-noise level of 100 has
been added to the light curve. This so constructed light curve shows
a clear asymmetry with different slopes for the ingress and egress,
different out-of-eclipse levels before and after eclipse as well as a
minimum at a phase $\not= 0$. These features are also apparent in the
superoutburst light curves of V2051~Oph (Fig.\ref{lcv_outburst}).

\begin{figure*}
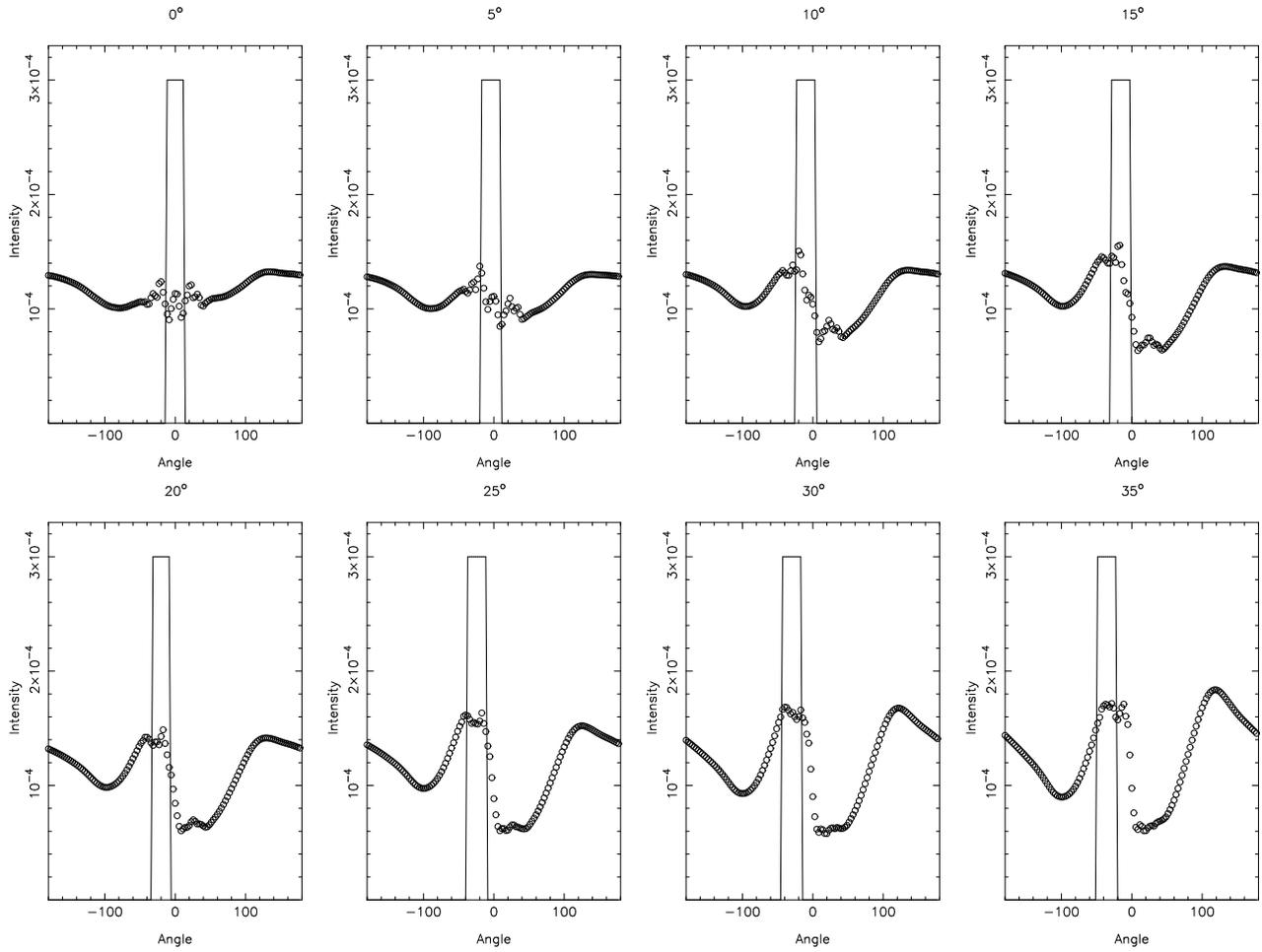

\hspace*{0.1cm}
\psfig{file=figA3a.ps,width=4cm}
\hspace*{0.1cm}
\psfig{file=figA3b.ps,width=4cm}
\hspace*{0.1cm}
\psfig{file=figA3c.ps,width=4cm}
\hspace*{0.1cm}
\psfig{file=figA3d.ps,width=4cm}

\vspace{1ex}
\hspace*{0.1cm}
\psfig{file=figA3e.ps,width=4cm}
\hspace*{0.1cm}
\psfig{file=figA3f.ps,width=4cm}
\hspace*{0.1cm}
\psfig{file=figA3g.ps,width=4cm}
\hspace*{0.1cm}
\psfig{file=figA3h.ps,width=4cm}
\caption{\small The reconstructed intensity distribution along the
disc ribbon for 8 different accretion disc orientations (across and
down) from $0^\circ$ to $35^\circ$ in $5^\circ$ steps as indicated
above each plot. The solid line indicates the position of the bright
area on the disc edge and the circles the reconstructed brightness
distribution.
\label{app_plotr}}
\end{figure*}

The application of the Eclipse Mapping algorithm to these light curves
lead to the reconstructed maps in Fig.~\ref{app_plotv} ({\em
right}). On the surface of the disc only the ``back'' part of the disc
(around an azimuth of $160^\circ$) shows some brightening, the front
part (around $-20^\circ$) seems to be unchanged. This is probably due
to a forshortening of the front part of the disc because of the
relatively large opening angle. In any case, the poor reconstruction
of the elliptical shape in this test case means that we cannot expect
to see an elliptical shape of the disc emission in Eclipse maps of
observed light curves.

However, on the edge of the disc (the ``ribbon'') a brightening can be
seen -- even if much fainter than in the original map. In
Fig.~\ref{app_plotr} we have plotted the light distribution along the
disc edge for reconstructions using eight different orientation angles
between $0^\circ$ and $35^\circ$. For larger angles the flaring will
be negligible.

Fig.~\ref{app_plotr} shows that in the reconstructions there is a large scale
brightness variation, often a peak near azimuth $0^\circ$ and for disc
orientation angles $>10^\circ$ a peak that coincides well
with the original position of the bright area on the disc
edge. Although the reconstructed brightness distribution for a disc
orientation angle of $0^\circ$ is not similar to the original one, it
is at least symmetric around azimuth $0^\circ$. For a disc
orientation angle of $5^\circ$ the reconstruction near azimuth
$0^\circ$ is rather noisy and should be recognizable in real data as a
disc with an orientation angle close to $0^\circ$.

This test shows that a peak in the brightness distribution along the
disc edge can be determined with a positional accurancy of $5^\circ$.

\end{appendix}
\end{document}